\documentclass{aa}
\usepackage{epsfig}
\usepackage{graphics}
\begin{document}
\thesaurus{11     
	     (11.01.2;  
	      11.14.1;  
	      11.19.3;  
	      11.19.7)}  
   \title{The FIR-Radio Correlation of Wolf-Rayet Galaxies and
   the Role of Star Formation in LINERs}

   \author{Ji, L., Chen, Y., Huang, J.H., Gu, Q.S., Lei, S.J.}

   \institute{Department of Astronomy, Nanjing University, Nanjing 210093, China}
	
   \titlerunning{Role of Star Formation in LINERs}
   \authorrunning{Ji et al.}
   \maketitle
  
   \begin{abstract}
      We find that a preliminary classification of LINERs' energetics may
   be made in terms of the FIR-radio correlation of Wolf-Rayet galaxies. 
   The AGN- or starburst-supported LINERs can be distinguished
   by their FIR-to-radio ratio,
   $Q\equiv L(1.4{\rm GHz})/$ $L(60\mu{\rm m})>$ or $<0.01$.
   It is interesting to note that almost all the LINERs with inner rings
   might be starburst-supported, indicating reduced AGN activities 
   compared with those
   of the AGN-supported ones. We also find that a shock-heating phase for the warm
   dust component might be important for some starbursts at the burst age of
   $\ge 10^{7}$ yr, with $Q<0.001$.

   \end{abstract}
%

\section{Introduction}

 Low ionization nuclear emission regions (LINERs, Heckamn 1980) are galaxies
with strong forbidden lines from low ionization states, compared with those
from higher ionization states, 
typically [O {\sc ii}] 3727 $>$ [O {\sc iii}] 5007. LINER phenomena are 
the most common activities of galaxies known  in the local universe
(Ho, Felippenko, \& Sargent 1997). If most LINERs are truely low luminosity
active galactic nuclei (LLAGN) (Ho 1998), it will be important  to 
understand the nature of AGN by investigating  the role of LINERs. However,
the origin of LINER phenomena is debated (Maoz et al 1998; Lawrence 1998;
    Ho 1998; and the references therein).
Maoz et al (1998) claimed  that the young stellar 
    population may provide enough ionizing photons for the observed spectra 
    of a significant fraction of LINERs, and
    it is also possible that LINERs may be a heterogeneous
    class. A good example is M81, one of the nearest bright LINERs.
    Ho (1998) stressed the incontrovertible,
    nonstellar nature of M81, though the stellar contribution 
    cannot be ruled out
    (Maoz et al. 1998). Recently, Hameed \& Devereux (1999) discussed the
    extended nuclear emission-line regions
    having a LINER spectrum in M81, and argued that 
    M81 is likely a composite object. They also proposed that
    shock heating or UV-photons
    from post asymptotic giant branch stars are probably responsible
    for the extended LINER emission.
    Further evidence for the ``composite'' 
    nature for M81 is provided by the ROSAT HRI data
    (Colbert \& Mushotzky 1999),
    in which the intensity ratio of the point-like component to the 
    extended one is about 3/2 (Colbert, private communication).

    It has been suggested (Condon et al. 1982) that the AGN activities
    can be distinguished from starbursts by using the FIR-radio 
    correlation, which is more significant for starbursts than for AGN. The
    weak FIR-radio correlation of
    Seyferts may imply the starburst (SB)-dominated bolometric luminosity for
    Seyferts (Forbes \& Norris 1998; also see, Norris et al. 1988).
 The hypothesis of a close connection between starbursts and Seyferts has
   been
proposed by Terlevich and his collaborators (Terlevich \& Melnick 1985; 
Terlevich et al 1992). They claim that the radio-quiet AGN are powered
by massive nuclear starburst in a metal-rich environment.
Recently, Heckman et al. (1997) discovered a powerful nuclear starburst
(in the Wolf-Rayet (WR) phase) in Seyfert galaxy Mrk~477.
The bolometric luminosity of Mrk~477 in the central region might be
dominated by the nuclear starburst. 
It was remarked by Maiolino et al. (1998) that Heckman et al's study would 
provide observational evidence partly supporting 
    Terlevich's hypothesis.

    Following these leads, 
    we have explored the energetics of LINERs by using
    the FIR-radio correlation for Wolf-Rayet (WR) galaxies, suggesting  
    different energy budgets of LINERs.

\section{FIR-radio correlation of WR galaxies}

    WR galaxies are extragalactic sources that exhibit broad emission lines
    characteristic of WR stars in their spectra (Conti 1991). Their 
 typical burst ages are $\sim 2$-$8$ Myr. The 50 WR galaxies that show a good
    FIR-radio correlation, as shown in Fig 1a, have
    detected flux at 1.4 GHz in NVSS Catalog (Condon et al. 1998) and
    60$\mu m$ in IRAS (Moshir et al. 1992) among 139 known sources
    (Schaerer et al. 1999).
    Such a correlation of WR galaxies is apparently nonlinear,
    with a regression coefficient of about 1.20, 
 as obtained before for other
    samples of galaxies (e.g. Fitt et al. 1988; Cox et al. 1988).

  This correlation can be understood in the framework 
of starburst phenomenon (Moorwood 1996; Lisenfeld et al. 1996). 
For simplicity, we take the stars/dust geometry
    to be close to a star-free shell of dust surrounding a central dust-free
    sphere of stars (Mas-Hesse \& Kunth 1999). 
 In this scenario, the radiation from the nuclear starburst (the optical,
ionizing and non-ionizing photons)
heats the dust grains, and the UV photons emitted from the nuclear massive
stars photoionize the gas.

    The radio flux at 1.4 GHz consists of thermal
    bremsstrahlung emission from photoionized gas
    and synchrotron radiation from supernova remnants. The luminosities of 
    the two radiation mechanisms are respectively given by Rubin (1968)
    \begin{equation}
       L^{\rm f-f}_{\rm T}(\nu) =
 1.59\times 10^{-32}\left(\frac{\nu}{\rm GHz}\right)^{-0.1} T^{0.45}_{4} N_{\rm UV}  \,\,{\rm W\,Hz^{-1}}
    \end{equation}
  where $T_{4}$ is electron temperature in units of
   10$^{4}$K and $N_{\rm UV}$ ionizing photons per second,
    and by Colina \& P\'{e}rez-Olea (1992)

   \begin{equation}
      L^{\rm SNR}_{\rm NT}(\nu) = 
  4.45\times 10^{22}\left(\frac{\nu}{\rm GHz}\right)^{-0.7} T_{\rm SNII} \,\,{\rm W\,Hz^{-1}}
   \end{equation}
   where $T_{\rm SNII}$ is the Type {\sc ii} supernova rate.

   The FIR radiation at 60$\mu$m is assumed to be composed of two
   parts: the
   warm dust
   component caused by the same starburst event, and the cool dust component
   outside the starburst region heated by the general interstellar radiation.
 Xu et al. (1994) modelled the contribution of cool and warm components
in FIR - radio correlation for late-type galaxies
   Several authors have virtually tried correcting or linearizing the
    FIR-radio
   correlation (Condon 1992; also see Fitt et al. 1988;
   Devereux \& Eales 1989).
   The luminosity of the FIR radiation is described by

   \begin{eqnarray}
      L_{\rm IR}(\nu) &=&
  4\pi B_{\nu}(T_{\rm d,w}) Q_{\rm abs}(\nu) \pi a^{2} N_{\rm d,w}\nonumber\\
     & & + 4\pi B_{\nu}(T_{\rm d,c}) Q_{\rm abs}(\nu) \pi a^{2} N_{\rm d,c}
   \end{eqnarray}
   where $T_{\rm d,w}$ and $T_{\rm d,c}$ are the warm 
   and cool dust temperatures,
   $N_{\rm d,w}$ and $N_{\rm d,c}$ the total number of warm and cool dust
    grains,
   $a$ is the average radius of dust grains, $Q_{\rm abs}(\nu)$ the
    absorption
   efficiency of dust grains, and $B_{\nu}(T_{\rm d})$ the Planck function.
   We adopt the ``astronomical silicate'' dust model (Draine \& Lee 1984),
   which is most likely suitable to starburst galaxies (Mas Hesse \& Kunth
   1999).

Assuming a ``steady-state'' case for the dust grains,
   the dust temperatures can be derived from the equilibrium 
   between dust absorption and dust emission,  
   \begin{equation}
    c\int_{0}^{\infty} U_{\lambda} Q_{\rm abs}(\lambda) d\lambda =
 4\pi \int_{0}^{\infty} B_{\lambda}(T_{\rm d}) Q_{\rm abs}(\lambda) d\lambda
  \end{equation}
   where $U_{\lambda}$ is the energy density of a diluted radiation field 
   that heats the dust, which is satisfied with
   \begin{equation}
   U_{\lambda} = \frac{4\pi}{c} B_{\lambda}(T_{\rm eff}) W
   \end{equation}
where $W$ is the dilution factor, and $T_{\rm eff}$ the equivalent 
effective temperature for the radiation field generated by starburst
activities.
 Using the $\lambda^{-1}$ dependence for $Q_{\rm abs}(\lambda)$, one
can yield the dust temperature from equa. (4):
$T_{\rm d}\sim T_{\rm eff} W^{\frac{1}{5} }$ (Spitzer 1978).
The FIR luminosity at 60$\mu$m 
can be obtained from equa. (3), scaling the value of
$Q_{\rm abs}(\lambda)$ to fit the Draine \& Lee (1984) model at 
$60\mu m$:
[$\lambda Q_{\rm abs}(\lambda)/a]_{60\mu m}~\sim~$2.5 .
  
At any given burst age, the evolutionary synthesis model, GISSEL95
(Bruzual \& Charlot 1996), is used to provide the relevant
quantities such as $N_{\rm UV}$, $T_{\rm SNII}$, and the bolometric
corrections for deriving the effective temperatures. Considering the
discussion by Mas Hesse \& Kunth (1999), we have assumed 50\% of 
$N_{\rm UV}$ are absorbed by dust.

We have estimated the possible values of the
dilution factor in various ways and adopt their average, 10$^{-14}$,
which is compatible with the usual interstellar value (Spitzer 1978).
The radiation transfer is not taken into account.
A dust-to-gas mass ratio $\sim1/100$ is assumed.
We also assume that the gas mass is comparable to
the star mass in the starburst region ($M_{\rm SB}$) (namely the gas-to-star
mass ratio is roughly unity). 
The total grain number is interpreted as $M_{\rm SB}/\rho_{d}a^{3}$ where
the density of the 'astronomical silicate' is adopted as
$\rho_{d}\sim 3{\rm g}\,{\rm cm}^{-3}$ (Draine \& Lee 1984).

For calculating the cool dust temperature, we assume that the cool dust
component may be heated by the general interstellar radiation field that
arises
from a past starburst event with a typical burst age $\ge$ 1 Gyr. The mass
of the cool component is a free parameter, and we try fixing its value,
 10$^{6} M_{\odot}$ or $5\times10^{4}M_{\odot}$,
 for any $M_{\rm SB}$. It means that the contribution
from the cool component is relatively significant for small burst strength
(small $M_{\rm SB}$), and relatively unimportant for
ultraluminous infrared galaxies (ULIGs; large $M_{\rm SB}$).
Generally, we have $T_{\rm d,c}\sim20$ K,
similar to the assuming cool dust temperature by Fitt et al. (1988).

To perform the calculations, we take
$M_{\rm SB}$ as an independent variable, which is in the range of
$10^{4.5}$-$10^{10}M_{\odot}$. The upper end of $M_{\rm SB}$ corresponds to
the case of ULIGs (e.g. Genzel et al.
1998). The stronger the starburst (i.e., the larger $M_{\rm SB}$),
the higher the FIR and radio luminosities.
With various adopted parameters (burst ages, dust-to-gas ratio, etc.),
we obtain {\em linear} FIR-radio correlations if taking only
the warm dust component into consideration, or a {\em nonlinear}
correlations if both the warm and cool dust components.
The model curves are plotted in Fig 1a, the model parameters are listed
in Table 1.

The solid line I in Fig 1a represents the linear part of our model prediction
 at the burst age of 6 Myr, in which the contribution of cool dust emission
   is neglected and thermal (bremsstrahlung) emission is dominant at 1.4 GHz.
 The dashed lines in Fig 1a illustrate the lower-right envelopes for models,
   in which the contribution of cool dust emission is taken 
   into account. For lines IIa and IIb, the cool dust mass is taken as
   $5\times10^{6}M_{\odot}$ and $10^{6}M_{\odot}$ 
 at the age of 3 Myr, respectively, and for line IIc, the cool dust mass is
   $5\times10^{4}M_{\odot}$ at the age of 6 Myr.
   As expected, counting the cool dust emission reproduces the nonlinear
   trend in the correlation lines.
   It is quite reasonable to see in Fig 1a that the cool dust component
   makes significant contribution in the case of small burst strength, while
   it is negligible compared with warm component for large burst strength.
   Satisfactorily, the majority of WR galaxies are located in a ``passage''
   escorted by the upper and lower envelopes in the diagram. 
   Reasonably, this passage may be considered to be 
   typical of the positions of the SB-dominated galaxies.

 In Fig 1b, we have added two prototypical starburst galaxies,
  M~82 and NGC~253,
   corresponding to a burst age of 10$^{7}$-10$^{8}$ yr. The dotted line Ia
 in Fig 1b indicates our model prediction (with a dust-to-gas ratio of 1/100)
   at the burst age of 2$\times$10$^{7}$ yr, which represents the upper end 
   of age for the supernova, set by GISSEL95.
   The non-thermal (synchrotron) radiation dominates at 
   1.4 GHz in this case. Considering the enrichment of the dust grains by 
   supernova explosions at this age, the dust-to-gas ratio can increase 
   by several times, up to
   an order of magnitude (Hirashita 1999), so it would be reasonable to 
   replace
   the dust-to-gas ratio of 1/100 with 1/20. As a result, the
   model curve will shift to a position indicated by the 
   dashed line Ib in Fig 1b.

   In order to fit the galaxies that exhibit ongoing star formation, 
   such as a transition object NGC~5194 (Heckman 1980; Larkin et al. 1998),
   we tried to add a shock wave (that may be related with the supernova
   explosions and/or outflowing winds from starburst) as additional mechanism
   for heating the dust,
   following Dwek (1986) and Contini et al. (1998). 
   The model curve containing a
  shock-heating phase is represented by the dot-dashed line Is in Fig 1b,
   which
   is below the lower border of the passage mentioned above. Here, we have 
   adopted a shock velocity $v_{\rm s}=200{\rm km}\,{\rm s}^{-1}$,
   a shock covering
   fraction 1/10, and a dust-to-gas ratio 1/20 at the age of
    $2\times10^{7}$ yr.
   It is worth noting that a strong near-infrared [Fe {\sc ii}] line
   has been observed in NGC~5194, and the shock
   excitation in supernova remnants is probably the mechanism responsible
   for this line (Larkin et al 1998). This excitation mechanism may be
   consistent with our consideration of shock-heating of dust in this
   galaxy, with $v_{\rm s}$ in order $\sim100\,{\rm km}\,{\rm s}^{-1}$.

\section{Energetics of LINERs}
\subsection{Classification of LINERs}
 Recently, more than a dozen of LINERs have been 
   studied with space facilities
   or large ground-based telescopes. Table 2 lists these sources with 
claims for their energetics, except for M81 because of the debate
 mentioned in
Sect. 1.
   Fig 2a indicates the positions where these LINERs are located
   in the FIR-radio
  diagram of WR galaxies. It is very instructive to see that the LINERs with
   SB-supported claims or with the existence of nuclear starburst, 
   in notation of SB in Table 2, have a similar distribution to 
   WR galaxies, while the AGN-supported LINERs or those with the
    existence of AGN
are distinctively located in the upper-left part of the FIR-radio diagram.
 The latter category of objects is found above
the upper border (solid line in Fig 2a) of our
models for starburst events.
 
Furthermore,  we investigate two cases of LINERs that are considered to have 
a composite nature, NGC6240 and M81. For NGC6240, Schulz et al. (1999) 
claimed,  based on ROSAT data, that both AGN and starburst contribute 
in roughly equal 
proportion to the energetics of this galaxy. In our Fig. 2a, both NGC6240 
and M81 are approximately located on the border of our model 
prediction. Due to the enrichment of dust and the influence of cool 
dust component as discussed in Sect. 2, the actual position of our model 
   prediction in Fig 2a might somewhat move to the right and curve up
    at small burst strength.

   Of particular importance to the AGN nature in LINERs is the detection of
   broad (FWHM $\sim$ a few thousand km $\rm s^{-1}$) permitted lines
    in these
   sources, which may arise from the broad-line regions (BLR). On the
   analogy of
   the nomenclature for Seyferts, Ho (1998) has designated those sources
   having visible BLR as LINER 1, and others as LINER 2.
 
It is enlightening to see that the most part of LINER 2's
 in Terashima (1999)
 have SB notations in our Table 2, while the majority of LINERs with 
   AGN notations in our Table 2 are listed as LINER 1's in Terashima (1999).  
 Now we have seen a {\it fact} that these LINER 1's are basically 
   {\it segregated} from LINER 2's in the FIR-radio diagram.
   This segregation confirms the early suggestion by Condon et al
   (1982) of distinguishing AGN from starburst by use of the FIR-radio 
   correlation.

 The majority of sources studied in this paper are at distances of 10 Mpc
or beyond, indicating that
the FIR-radio diagrams shown in our figures are basically
referred to the global 
   properties of galaxies, due to the large
resolution/apertures used in the NVSS and the IRAS.
   Nevertheless, what we have seen in the segregation is the pairing of
   the observed
   SB-supported LINERs to modeled SB-dominated passage,
 {\it not} in a
   cross-pair of the AGN- to SB-dominated passage. 
   It would be hard to understand that the
   segregation could be just caused by chance.

 The studies on LINERs in the
IRAS 1-Jy (f(60$\mu{\rm m}) >$ 1 Jy) sample of ultraluminous
infrared galaxies
   (LINER ULIGs) by Veilleux et al. (1999) may shed light on
    the above fact.
   They conclude that ``there is no convincing
   optical or infrared evidence for an AGN in LINER ULIGs'', and
     ``the main
   source of energy in these LINERs is a starburst rather than an AGN.''
   We have put these LINER ULIGs in the FIR-radio diagram
   in Fig 2a by symbols of crosses. Their positions are in the
   SB-dominated passage.  Recent studies 
   (see, Veilleux et al 1999 and the references therein) strongly
   suggest that the overwhelming part of the bolometric luminosity of ULIGs 
   stems from the inner kpc regions, indicating that the large
resolution/apertures used in the NVSS and the 
IRAS would not make any obvious change 
   in the FIR-radio correlation for LINER ULIGs. In fact, these studies are
   consistent with the early work by Kennicutt \& Kent (1983), which 
   demonstrated that in
   the case of EW(H$\alpha$) $\ge$ 10\AA , suitable to LINER ULIGs,
   the H$\alpha$ emission observed with a slit is comparable to that 
   obtained using large apertures.

   Similar analyses of early-type galaxies (Walsh et al 1989)
   suggest no obvious variance
   in the FIR-radio correlation with the sizes of galaxies.
   Radio observations of a sample of ellipticals and S0s
    (Fabbiano et al 1987)
   have not found evidence for the extended disk emission, confirming the
   suggestion that the radiation is from nuclear ``starburst''
    instead of extended
   disk sources (Dressel 1988). These investigations provide
   a sound explanation
   for the statistical study by Walsh et al. (1989), implying that the
   location of the AGN-supported LINERs (the majority of which are E/S0
   sources) in the upper-left part of FIR-radio diagram (Fig 2a) may not be 
   significantly influenced by the size of apertures used.

On the other hand,
NGC~6500, a spiral far above the border of the starburst events in Fig 2a,
 has
   EW(H$\alpha$) = 27\AA~ (Ho et al 1997), indicating little effect
    of the aperture sizes
on the FIR-radio correlation, according to Kennicutt \& Kent (1983).
   The same argument of EW(H$\alpha$) $\ge$ 10\AA~ holds for NGC~404,
    one nearby
 galaxy, and four other LINERs selected from the Pico dos Dias Survey (PDS)
 (Coziol et al. 1998). Three of the PDS LINERs are classified as transition 
   sources by Coziol et al. (1998), including NGC~3310, designated as
   starburst in V\'{e}ron Catalog, that is classified as a transition source, SB/LINER.

 The remaining sources, NGC~4736, NGC~5055, and NGC~7217, are
  three LINERs with
  EW(H$\alpha$) $<$ 10\AA.
  For example, the EW(H$\alpha$) of NGC~7217 obtained with a slit and 
  a large aperture are about 3\AA~ and 6\AA~
  (Ho et al 1997; Kennicutt \& Kent 1983), respectively.
  After correcting the aperture effect over this source, the
  position of NGC~7217 in Fig 2a may move a bit down- and leftward
 in the FIR-radio diagram,
 remaining in the SB-dominated passage. 
 The same argument would be applicable to 
 other sources with EW(H$\alpha$) $<$ 10\AA.

   From the above discussion, one can see that the aperture
    effects
   may not significantly change the situation of segregation for different
   types of LINERs shown in Fig 2a. Therefore, as
   suggested by Condon et al (1982),
  the FIR-radio correlation may provide a {\em preliminary classification} 
 of LINERs according to their locations in the diagram as we described
 above. In other words, one may classify LINERs
in terms of their FIR-to-radio ratio:
 $Q\equiv L(1.4{\rm GHz})/L(60\mu{\rm m)}$:
   one has $Q>0.01$ for the AGN-supported LINERs and $Q<0.01$ for the
    SB-supported ones.

  In Table 3 we list part of
   LINERs extracted from V\'{e}ron Catalog
    (V\'{e}ron-Cetty \& V\'{e}ron 1996)
   that have detected fluxes at 1.4 GHz and 60${\mu}$m,
   and the preliminary classification of their
   energetics is given in column 6.
    Their distributions in the FIR-radio diagram
   are shown in Fig 2b. 

  As further evidence our classification the types of LINERs, 
  we mention the new results of Alonso-Herrero et al. (1999). The different 
  types of LINERs identified by these authors are consistent with our 
   predictions for those common ones, as listed in our Table 3.
   For example,  the claimed SB-dominated LINERs in their paper,
    NGC~3504, NGC~3367, NGC~4569,
   NGC~4826, and NGC~7743 are all located in our SB-dominated passage, and
   the AGN-dominated LINER NGC~2639 (Alonso-Herrero et al 1999) is designated
   to be an AGN-supported LINER in Table 3.
   Further observations of these sources are certainly needed, especially for
   the LINERs located near the boundary of the starburst events
    shown with solid
   line in Fig 2, which might be composite like NGC~6240.

\subsection{LINERs with inner rings}
      The effects of rings or bars on Seyfert activities have been
   studied since the early work in the 1980's (e.g. Simkin, Su, \& Schwarz 
 1980; Arsenault 1989). Recent studies show, however, that the frequency of
barred systems is the same in Seyferts and in normal spirals
(McLeod
\& Rieke 1995; Ho, Filippenko, \& Sargent 1997; Mulchaey \& Regan 1997). A
   latest study on the morphology of the 12${\mu}$m Seyfert Sample (Hunt
   et al 1999) indicates that LINERs have higher rates of inner rings than
   normal galaxies.

   One striking feature in Fig 2a and 2b is that almost all the LINERs 
   having inner rings are located in the SB-dominated passage. 
 This strong morphological tendency in the LINER sample should have important 
   implication in starburst-AGN connection. Our preliminary analysis shows 
   that the AGN activities of these LINERs are lower than the AGN-supported
   LINERs located in the upper-left part of the
   FIR-radio diagram. The reason for the reduced activities might be caused
  by the reduced fueling gas to the central black holes. We will discuss this
topic in a separate paper (Lei et al. 1999),
 along with the morphological study
   of Seyferts.

\begin{acknowledgements}
We wish to thank Dan Maoz for valuable comments that helped 
improve our manuscript. The anonymous referee is thanked for his/her critical
comments and kind help to improve our English presentation.
We are very grateful to Wei Zheng for his careful reading of the manuscript.
      This work is supported by grants from the NSF of China,
      and grants from the Ascent Project of the State Scientific Commission 
      of China.
\end{acknowledgements}

\newpage 
    \begin{table*}
      \caption[]{Model parameters}
      \[
	\begin{array}{lccccccc}
	  \hline
	   &   & & Fig. 1a &   &  &Fig. 1b & \\
	  \hline
	  & I&IIa & IIb &IIc& Ia &Ib & Is\\
	  {\rm IMF}& && & {\rm Salpeter} & & &\\
	  {\rm mass\hspace{1mm}range}& && 0.1 M_{\odot} &\sim& 125 M_{\odot} & &\\
	  {\rm metalicity}& && & Z_{\odot}& & &\\
	  T_{\rm e}({\rm K})& && & 10^{4} & & &\\
	  M_{\rm dust}/M_{\rm gas} & 1/100 & 1/100 &1/100& 1/100& 1/100 & 1/20 & 1/20\\
	  {\rm age(Myr)}& 6 & 3 & 3&6 & 20 & 20 & 20\\
	  M_{\rm cool}(M_{\odot})&- & 5\times 10^{6} &10^{6} & 5\times10^{4} & -&-& -\\
	  n_{\rm e}({\rm cm}^{-3}) & -& -& -& -& -&-& 300\\
	  v_{\rm s}({\rm km\hspace{1mm}s}^{-1}) & - & -& -& -& -&-& 200\\
	  T_{\rm d,w}({\rm K})^{\star} & 31 & 43 & 43 &31& 27&27 & 46^{\rm s},27 \\
	  T_{\rm d,c}({\rm K})^{\star} & - & 20 & 20 &20& -&-&- \\
	  \hline
	\end{array}
     \]
\begin{list}{}{}
\item{$\star$}  derived values
\item{s}  temperatures of the warm dust in shock region
\end{list}
    \end{table*}
\newpage 
   \begin{table*}
      \caption[]{LINERs with claims}
      \[
	\begin{array}{llccccl}
	   \hline
	   {\rm Name} &{\rm Type} &\log(L_{60\mu {\rm m}})&\log(L_{\rm 1.4GHz})&D &{\rm Energetics} &{\rm References}\\
	  &&(10^{29}{\rm erg}\,{\rm s}^{-1} {\rm Hz}^{-1})&(10^{26}{\rm erg}\,{\rm s}^{-1} {\rm Hz}^{-1})&({\rm Mpc})&&\\
	   \hline
	   {\rm M~84}   &.E.1...&  0.3   &3.7    &   16.8 &{\rm AGN} & {\rm H98}\\
	   {\rm M~87}     &CE.0...& 0.1 &6.2& 16.8 & {\rm AGN}& {\rm H98,Col98}\\
	   {\rm NGC~1052} &.E.4...&  0.6 &3.5& 17.8 &  {\rm AGN} & {\rm B98}\\
	   {\rm NGC~3718}  &.SBS1P.&    0.3    &1.7  &  17.0  &{\rm AGN}& {\rm H98}\\
	   {\rm NGC~3998} &.LAR0\$.&  0.5        &2.7& 21.6 & {\rm AGN} &{\rm L98}\\
	   {\rm NGC~4203} &.LX.-*.&  -0.1 &0.9 & 9.7 & {\rm AGN}&{\rm H98,I98}\\
	   {\rm NGC~4278} &.E.1+..*& -0.1        &2.6& 9.7 & {\rm AGN}&{\rm F98}\\
	   {\rm NGC~4594} &.SAS1./&   1.3&   2.6 &20.0 & {\rm AGN}&{\rm M98,N97}\\
	   {\rm NGC~4579} &.SXT3..&   1.3&   2.3 &16.8   & {\rm AGN}&{\rm M98,L98}\\
	   {\rm NGC~6500} &.SA.2*.&  1.1 &3.5& 39.7 & {\rm AGN} &  {\rm F98,M98}\\
	   {\rm M~81}     &.L.....&  -0.5&  0.3 &1.4  & {\rm Composite}&{\rm H98,M98,Col98}\\
	   {\rm NGC~6240} &.I.0.*P&  3.4&   4.7 & 98.1 &  {\rm Composite}&{\rm S98}\\
	   {\rm NGC~404} &.LAS-*.&  -0.8&   -0.6& 2.4 &   {\rm SB} & {\rm M98}\\
	   {\rm NGC~5055}& .SAT4..&  1.4&   1.4  & 7.2 &   {\rm SB}& {\rm M98}\\
	   {\rm NGC~4736}&RSAR2..&   1.2&   1.0  & 4.3 &   {\rm SB}& {\rm L98}\\
	   {\rm NGC~5194}&.SAS4P.&   1.8&   1.9  & 7.7 &   {\rm SB}& {\rm L98}\\
	   {\rm NGC~7217}&RSAR2..&   1.2&   1.4  & 16.0 &   {\rm SB}&{\rm L98}\\
	   {\rm NGC~4818}    &.SXT2P*&  1.7  & 2.0     &13.5 &{\rm SB}& {\rm Coz98}\\
	   {\rm NGC~3310}    &.SXR4P.&  2.2  & 2.9    &18.7 &{\rm SB}&  {\rm Coz98}\\
	   {\rm NGC~1050}    &PSBS1..&  2.3   & 3.0     &53.4  &{\rm SB}& {\rm Coz98}\\
	   {\rm UGC~03157}   &.SB?...&   2.2  &  3.0    & 61.4   &{\rm SB}&  {\rm Coz98}\\
	   {\rm Arp~220}  &.S?....&  3.9&   4.3 & 73.7 &  {\rm SB} &{\rm G98} \\
	   \hline
	\end{array}
      \]
   \end{table*}

\newpage 
   \begin{table}
     \caption{LINERs from V\'{e}ron Catalog}
     \[
	\begin{array}{llcccr}
	   \hline
	   {\rm Name} & {\rm Type}&\log(L_{60\mu\rm{m}})& \log(L_{\rm 1.4GHz})& D &{\rm Energetics} \\
	   & &(10^{29}{\rm erg}\,{\rm s}^{-1}{\rm Hz}^{-1})& (10^{26}{\rm erg}\,{\rm s}^{-1}{\rm Hz}^{-1})& ({\rm Mpc}) &\\
	   \hline
	   {\rm IC~1459} &.E.....&   0.5 &  3.9&  22.4& {\rm AGN}  \\
	   {\rm IC~1481} &.S?....&   2.1 &  3.5 & 83.5& {\rm AGN}  \\
	   {\rm NGC~2639}&RSAR1*\$&   1.7 &  3.4&  44.7& {\rm AGN}  \\
	   {\rm NGC~2655}&.SXS0..&   1.1 &  2.9&  24.4& {\rm AGN}  \\
	   {\rm NGC~2911}&.LAS.*P&   0.6 &  3.0&  40.8& {\rm AGN}  \\
	   {\rm NGC~3312}&.SAS3P\$&   1.2 &  2.7&  35.4& {\rm AGN}  \\
	   {\rm NGC~4036}&.L..-..& 0.6 & 1.9 & 24.6 &{\rm AGN} \\
	   {\rm NGC~4438} &.SAS0P*& 1.2 & 1.3 &  16.8& {\rm AGN}  \\
	   {\rm NGC~5675} &.S?....&  1.3 &  3.6 & 55.4& {\rm AGN}  \\
	   {\rm NGC~7135} &.LA.-P.&  0.4 &  1.8 & 29.9& {\rm AGN}  \\
	   {\rm Mrk~266NE} &.P.....&3.1&   4.2& 112.7& {\rm AGN}  \\
	   {\rm Mrk~984} &.S..8*.&  2.4  & 4.3 &191.5& {\rm AGN}  \\
	   {\rm NGC~660} &.SBS1P.&  2.1  & 2.7 & 11.8&  {\rm SB}  \\
	   {\rm NGC~1961}&.SXT5..&   2.4 &  3.3&  54.1& {\rm SB}  \\
	   {\rm NGC~2841}&.SAR3*.&   1.0 &  1.0&   12.0& {\rm SB}  \\
	   {\rm NGC~3079}&.SBS5./&   2.4 &  3.4&  20.4& {\rm SB}  \\
	   {\rm NGC~3226}&.E.2.*P&   1.8 &  2.8&  23.4& {\rm SB}  \\
	   {\rm NGC~3367} &.SBT5..&  2.1 &  3.0 & 39.3& {\rm SB}  \\
	   {\rm NGC~3561B}&.SAR1P.&   2.8&   3.6& 113.7& {\rm SB}  \\
	   {\rm NGC~3994} &.SAR5P\$&  2.3 &  2.7 & 41.9& {\rm SB}  \\
	   {\rm NGC~4102} &.SXS3\$.&  2.2 &  2.9 & 17.0& {\rm SB}  \\
	   {\rm NGC~4192} &.SXS2..& 1.5 &  1.8 &  16.8& {\rm SB}  \\
	   {\rm NGC~5005} &.SXT4..&  2.1 &  2.7 & 21.3& {\rm SB}  \\
	   {\rm NGC~5371} &.SXT4..&  2.0 &  2.1 & 37.8& {\rm SB}  \\
	   {\rm NGC~5851} &.S?....&  1.9 &  2.6 & 87.4& {\rm SB}  \\
	   {\rm NGC~5921} &.SBR4..&  1.5 &  1.6 & 25.2& {\rm SB}  \\
	   {\rm NGC~5929} &.S..2*P&  2.2 &  3.1 & 35.8& {\rm SB}  \\
	   {\rm NGC~5953} &.SA.1*P&  2.0 &  2.8 & 27.1& {\rm SB}  \\
	   {\rm Mrk~313} &PLBS0*.&  1.8 &  2.2 & 28.5& {\rm SB}  \\
	   {\rm Mrk~700} &.S?....&  2.7  & 3.5& 135.6& {\rm SB}  \\
	   {\rm Mrk~848B}&.L...\$P&  3.5 &  4.2& 162.4& {\rm SB}  \\
	   {\rm IC9~10}  &.S?....&   2.8  & 3.7 &108.6& {\rm SB}  \\
	   {\rm ESO~568-G11}&.SXS4*P&1.9&   2.9& 118.7& {\rm SB}  \\
	   {\rm UGC~10082S} &.SB?...&2.3&   3.1& 141.1& {\rm SB}  \\
	    \hline
	 \end{array}
      \]
   \end{table}
%
\newpage 
   \begin{figure*} 
   \begin{center}\mbox{\epsfxsize=8cm \epsfbox{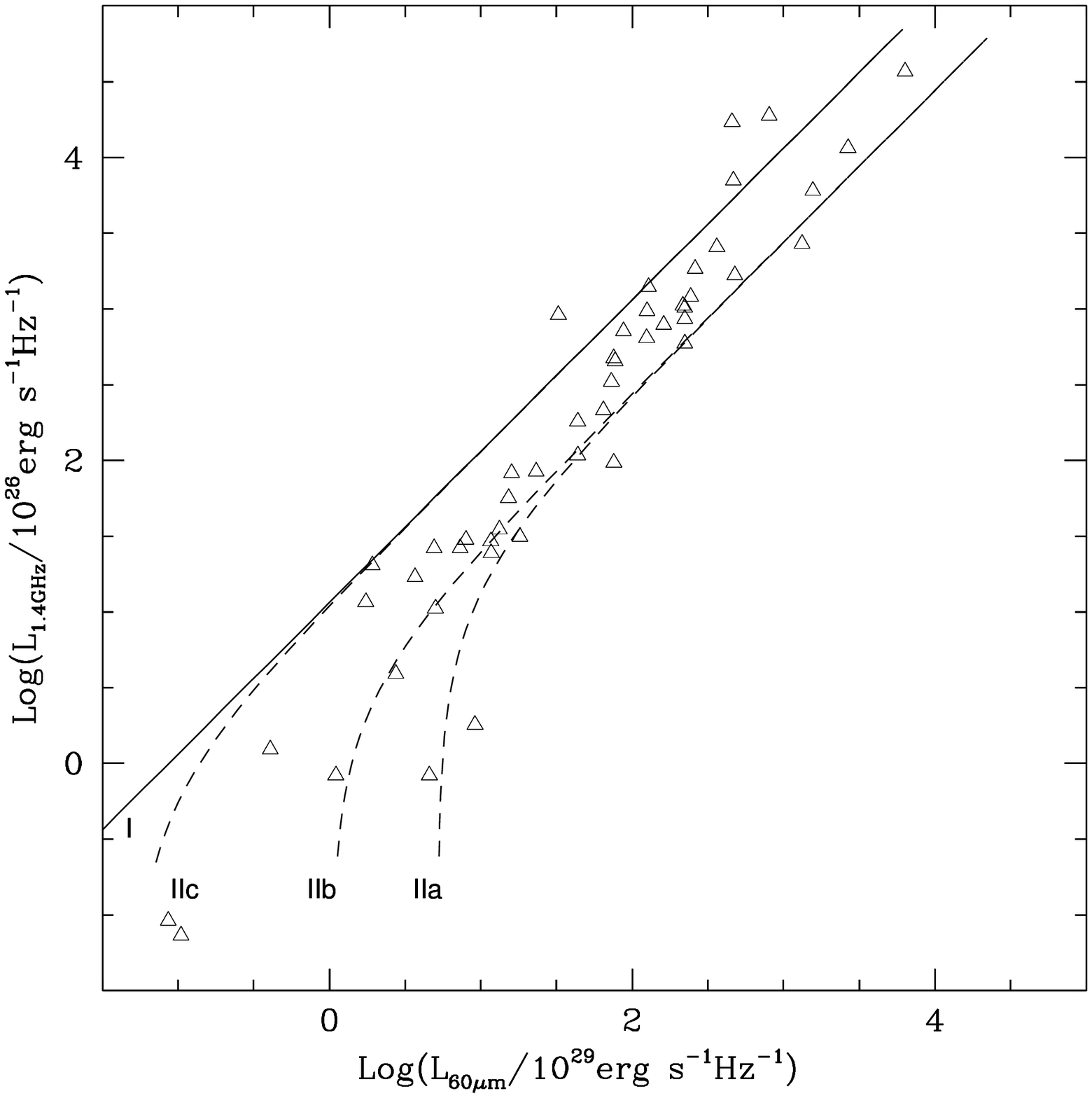}}\end{center}
   \begin{center}\mbox{\epsfxsize=8cm \epsfbox{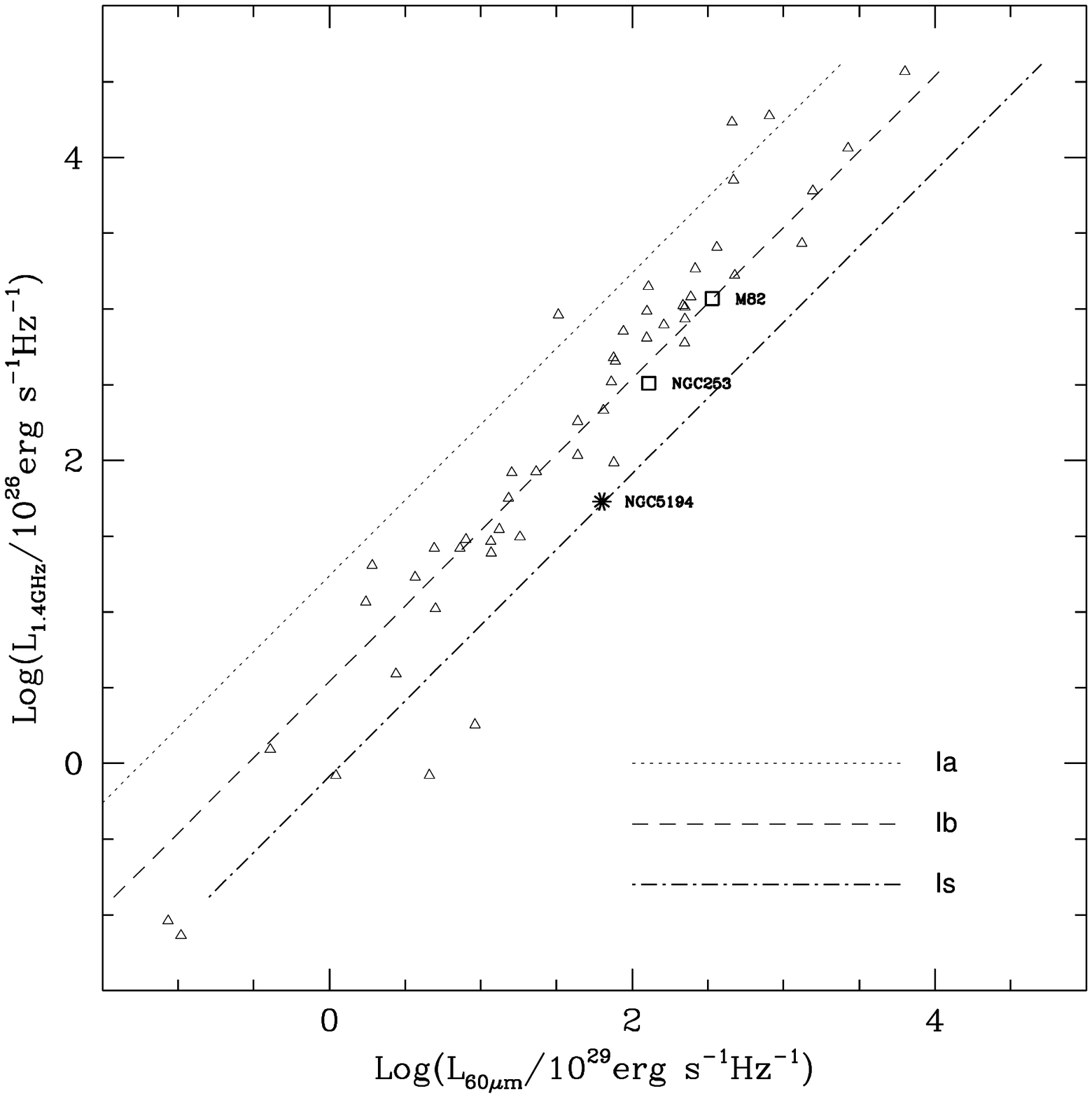}}\end{center}
      \caption[]{
 a) FIR-radio correlation of WR galaxies, denoted by open triangles.
    Model predictions are shown by lines: solid line for warm dust 
    component at the burst age of 6 Myr; dashed lines for models containing
    both warm and cool dust components, see the text for details.
 
 b) Same correlation as Fig 1a, added several sources at
    the burst age of 10$^{7}$ - 10$^{8}$ yr. Model predictions for the age of 
    2$\times$10$^{7}$ yr are indicated by lines: dotted line for 
    dust-to-gas ratio of 1/100; dashed line for dust-to-gas 
    ratio of 1/20; dot-dashed line for containing shock-heating phase but with
    same parameters as those for dashed line, see text for further details.
	}
   \label{cext}
   \end{figure*}
%
\newpage 
   \begin{figure*} 
   \begin{center}\mbox{\epsfxsize=8cm \epsfbox{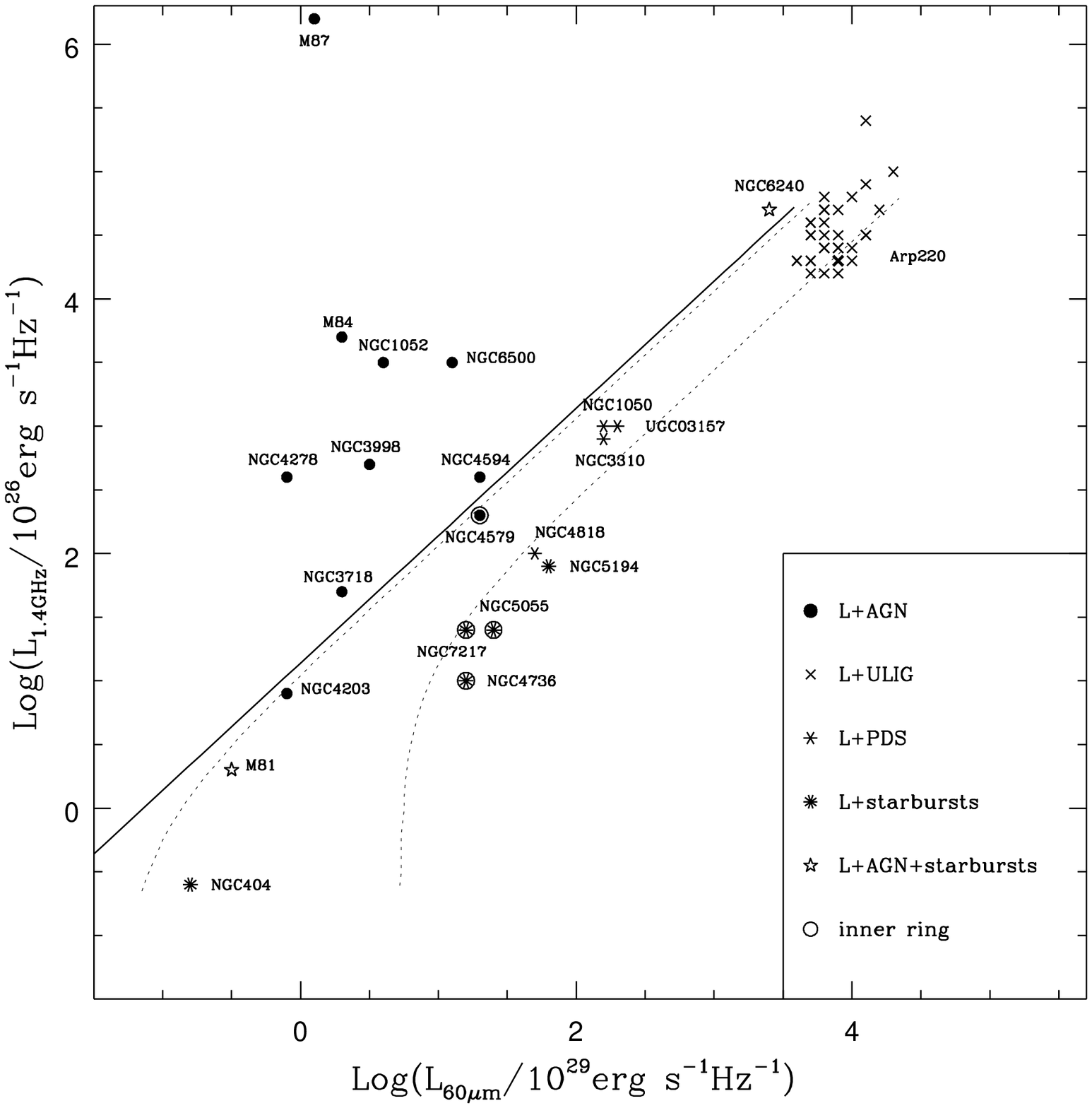}}\end{center}
   \begin{center}\mbox{\epsfxsize=8cm \epsfbox{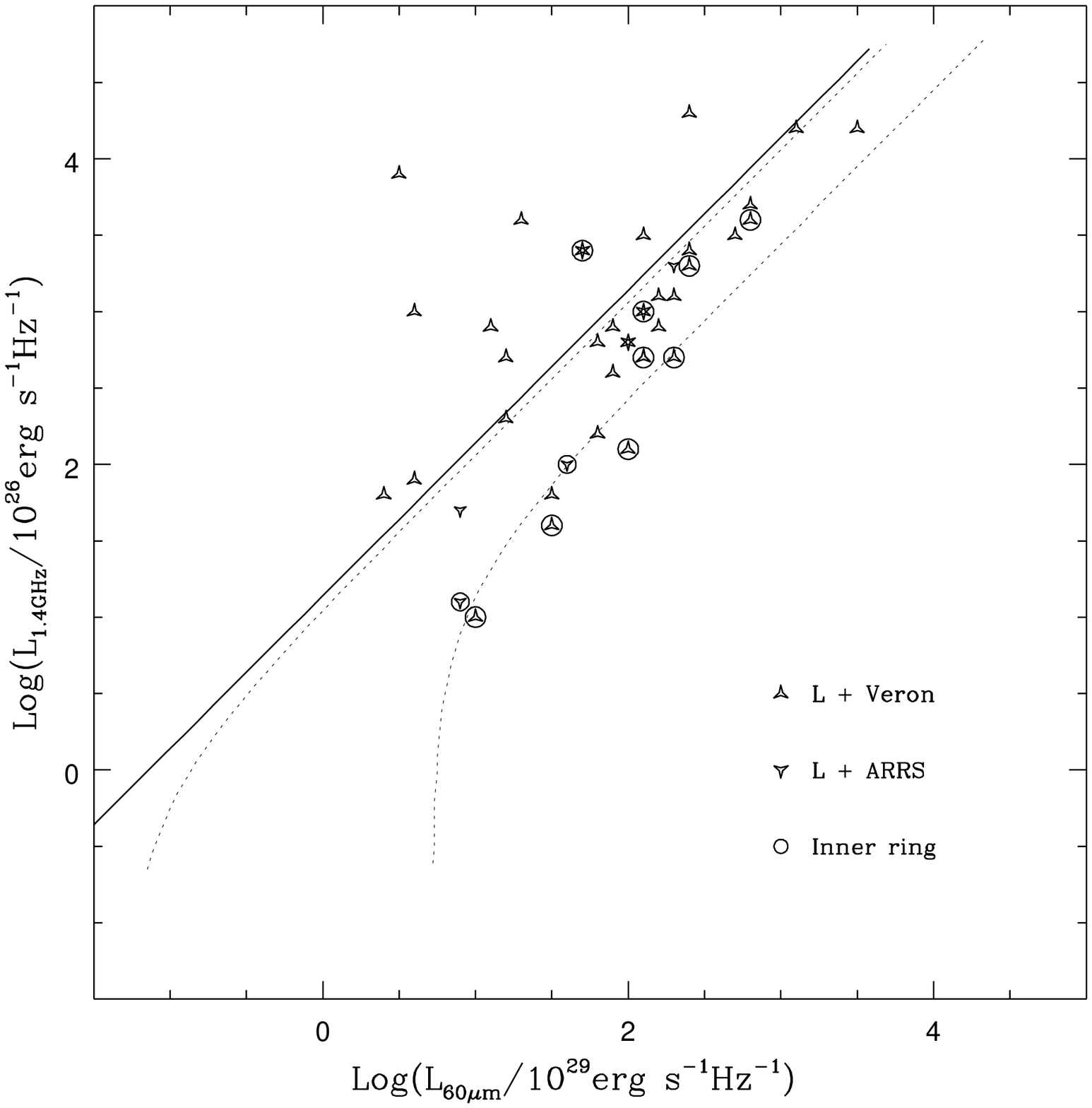}}\end{center}
      \caption[]{
 a) Same correlation diagram as Fig 1a, superposed by
    LINERs with claims of SB-supported or AGN-supported ones,  selected
    from literature. Dotted lines cover the area where the majority of WR
    galaxies are located. Solid line illustrates the prediction of starburst
    event at age of 10$^{7}$ yr. The notation of L in the lower-right box
    denotes LINERs. The notation of (L$+$AGN$+$Starburst) for M81 is a symbol 
    of ``composite'' for this source, see the text for details. 

 b) Same as Fig 2a, but the superposed LINERs are
    unclassified ones by symbols of starred triangles,
    extracted from V\'{e}ron Catalog, and some
    identified types of LINERs by symbols of inverse-starred triangles,
  selected from Alonso-Herrero et al. (1999). The symbols of hexa-stars are
    referred to the common ones between the two sets.
    }
    \label{cext}
   \end{figure*}
%
\end{document}